\documentclass[twocolumn,pra,aps,showpacs,groupedaddress,byrevtex]{revtex4}
\usepackage{graphics,ulem}
\usepackage{hyperref}
\usepackage{amsmath}

\newcommand{\tobj}{T_\mathrm{ obj}}
\newcommand{\ttobj}{{\cal F}\left\{T_\mathrm{ obj}\right\}}
\newcommand{\ttobjHBT}{{\cal F}\left\{|T_\mathrm{ obj}|^2\right\}}
\newcommand{\q}{\mathbf{q}}
\newcommand{\x}{\mathbf{x}}
\newcommand{\xp}{\mathbf{x}'}

\newcommand{\xv}{\mathbf{x}}
\newcommand{\rv}{\mathbf{r}}
\newcommand{\xiv}{\mathbf{x}'}
\newcommand{\im}{i}
\newcommand{\snf}{\Delta x_{\rm n}}
\newcommand{\sff}{\Delta x_{\rm f}}
\renewcommand{\em}{\it}%

\begin{document}
\title{Coherent imaging of a pure phase object with classical
  incoherent light}

\author{M. Bache} 
\email[Corresponding author: ]{bache@com.dtu.dk}
\altaffiliation{Present address: COM$\bullet$DTU, 
  Technical University of Denmark, DK-2800 Lyngby, Denmark}
\affiliation{CNR-INFM-CNISM, Dipartimento di Fisica e Matematica, Universit{\`a} dell'Insubria,
  Via Valleggio 11, 22100 Como, Italy}
\author{D. Magatti}
\affiliation{CNR-INFM-CNISM, Dipartimento di Fisica e Matematica, Universit{\`a} dell'Insubria,
  Via Valleggio 11, 22100 Como, Italy}
\author{A. Gatti}
\affiliation{CNR-INFM-CNISM, Dipartimento di Fisica e Matematica, Universit{\`a} dell'Insubria,
  Via Valleggio 11, 22100 Como, Italy}
\author{F. Ferri}
\affiliation{CNR-INFM-CNISM, Dipartimento di Fisica e Matematica, Universit{\`a} dell'Insubria,
  Via Valleggio 11, 22100 Como, Italy}
\author{E. Brambilla}
\affiliation{CNR-INFM-CNISM, Dipartimento di Fisica e Matematica, Universit{\`a} dell'Insubria,
  Via Valleggio 11, 22100 Como, Italy}
\author{L.A. Lugiato} 
\affiliation{CNR-INFM-CNISM, Dipartimento di Fisica e Matematica, Universit{\`a} dell'Insubria,
  Via Valleggio 11, 22100 Como, Italy}
 \date{April 6, 2006}
\begin{abstract}
  By using the ghost imaging technique, we experimentally demonstrate
  the reconstruction of the diffraction pattern of a {\em pure phase}
  object by using the classical correlation of incoherent thermal
  light split on a beam splitter. The results once again underline
  that entanglement is not a necessary feature of ghost imaging.  The
  light we use is spatially highly incoherent with respect to the
  object ($\approx 2 \mu$m speckle size) and is produced by a
  pseudo-thermal source relying on the principle of near-field
  scattering.  We show that in these conditions no information on the
  phase object can be retrieved by only measuring the light that
  passed through it, neither in a direct measurement nor in a
  Hanbury Brown-Twiss (HBT) scheme. In general, we show a remarkable
  complementarity between ghost imaging and the HBT scheme when
  dealing with a phase object.
\end{abstract}

\maketitle

\section{Introduction}
\label{sec:Introduction}

Ghost imaging is a technique which allows to perform coherent imaging
with incoherent light by exploiting the spatial correlation between
two beams created by, e.g., parametric down conversion (PDC)
\cite{klyshko:1988,belinskii:1994,strekalov:1995,pittman:1995,
  ribeiro:1994,ribeiro:1999,saleh:2000,abouraddy:prl-josab,
  abouraddy:2004,bennink:2002a,bennink:2004,
  gatti:2003,thermal,gatti:2004a,bache:2004,bache:2004a,ferri:2004,
  brambilla:2004a,bache:2005,gatti:2006,lugiato:2004,gatti:2005a,bache:2005b,cheng:2004,
wang:2004, cai:2004d, valencia:2004,zhai:2005}.  Each of the
correlated beams are sent through a distinct imaging system,
traditionally called the test and the reference arm.  In the test arm
an object is placed and information about the object is recreated from
the spatial correlation function between the test and reference arm.

One of the most striking features of the ghost imaging technique is
that since the two beams are spatially incoherent no phase sensitive
information about an object can be extracted by observing a single
beam only. This means that the diffraction pattern of an object that
substantially alters the phase of the incoming light cannot be
observed in any way in the test arm. Nonetheless, because of the
mutual spatial correlation between the two beams, we will show that
the diffraction pattern may be reconstructed through the spatial
correlation between the two beams. In other words, despite the beams
being incoherent, the coherence {\em between them} allows to
perform coherent imaging: the scheme is therefore capable of doing
coherent imaging with incoherent light.
\par
On the other hand, the diffraction pattern of an object that only
alters the amplitude of the light, such as a Young's double slit, can
be extracted from the intensity distribution of the object arm only,
even when the object is illuminated with a spatially incoherent beam.
One may do so by measuring the autocorrelation of the transmitted
field as observed in the far zone of the object. This can be
conveniently done by using, the Hanbury Brown-Twiss (HBT) scheme
\cite{hanburybrown:1956}. Thus, the ghost imaging technique is not
required in this case, and one may stick to using a single beam only.
\par
It is therefore interesting to look beyond the case of an
amplitude-only object.  As an extreme case, in this work we want to
observe the diffraction pattern of a {\em pure phase object}, an
object that only alters phase information
\cite{abouraddy:2004,thermal,bache:2004}.
\par
Initially the possibility of performing ghost imaging was ascribed to
the presence of spatial entanglement between the two arms
\cite{strekalov:1995,pittman:1995,ribeiro:1994,abouraddy:prl-josab}.
Lately this view has been challenged by many groups
\cite{bennink:2002a,gatti:2003,thermal,cheng:2004,ferri:2004,gatti:2006,
  bache:2005,brambilla:2004a,lugiato:2004,gatti:2005a,bache:2005b,
  bennink:2004,valencia:2004,wang:2004,cai:2004d,zhai:2005} (to cite but a few).
Our group in particular has produced numerous publications, showing
both theoretically \cite{thermal} and experimentally
\cite{ferri:2004,gatti:2006,bache:2005} that basically all features
offered by entangled ghost imaging can be mimicked by using a proper
scheme that exploits classically correlated beams: the correlation
between the beams is in this scheme created by dividing an incoherent
pseudo-thermal speckle beam on a beam splitter (BS).  The two outgoing
beams are then still incoherent on their own but since they are
(classical) copies one of each other, they have a high mutual spatial
coherence. We showed that the only feature that cannot be mimicked by
classical correlation is the $100\%$ visibility of the information,
that can be in principle achieved only with entangled photons;
however, in the classical case the visibility is still good enough to
effectively reconstruct the information
\cite{thermal,ferri:2004,brambilla:2004a,lugiato:2004,gatti:2005a,gatti:2006,
bache:2005,bache:2005b}.
An important outcome of our analyses is that the entangled ghost
imaging and our classical ghost imaging have the common feature of
providing coherent imaging using incoherent light
\cite{thermal}. Thus, both schemes should be able to
reconstruct the diffraction pattern of any object, altering amplitude
as well as phase \cite{thermal}. We experimentally
confirmed this prediction in the case of an amplitude object
\cite{ferri:2004}.
\par
An experiment of Abouraddy {\em et al.} \cite{abouraddy:2004}
demonstrated the reconstruction of the ghost diffraction pattern of a
pure phase object using the entangled photon pairs produced by
spontaneous parametric down-conversion, which represents an optical
field that lacks second-order spatial coherence but is endowed with
higher order spatial coherence.  The introduction of
Ref.~\cite{abouraddy:2004} claims that our ghost-diffraction experiment
with classical thermal light reported in Ref.~\cite{ferri:2004} could
be performed only because effectively the thermal light was endowed
with second-order spatial coherence. This might suggest that coherent
imaging with spatially incoherent light is not possible with split
thermal light.  This interpretation of our experiment
\cite{ferri:2004} was not correct, because the light we used there was
indeed spatially incoherent to a high degree. The viewpoint
of Ref.~\cite{abouraddy:2004} was challenged in
Refs.~\cite{bache:2005,gatti:2006}, where not only we experimentally
demonstrated again the ghost diffraction of an amplitude object with
classical incoherent light, but also showed that the spatial
incoherence of the light is a necessary ingredient to carry out the
task.  In this work we will finally demonstrate experimentally and
theoretically that the claim of Ref.~\cite{abouraddy:2004} is not correct
also in the case of a pure phase object.
\par
We will show that the diffraction pattern of a commonly used pure
phase object, a transmission grating beam splitter (TGBS), can be
reconstructed from the classical correlations between two highly
spatially incoherent beams generated by splitting a speckle beam on a
BS.  In order to render the light impinging on the object incoherent
with respect to the object, we have to produce speckles of size on the
order of $2.0~\mu$m, and we achieve it through the so-called
near-field scattering technique \cite{giglio:2000,giglio:2004}.
Incidentally we remark that such a small speckles size, which imply a
spatial resolution on the order of $2.0~\mu$m, (see, e.g.,
Ref.~\cite{ferri:2004}) has no precedent to our knowledge in ghost imaging
schemes, either with thermal or entangled beams.

With such a degree of spatial incoherence, we will verify that no
information about the phase object diffraction pattern is present in
the test arm, neither in the far zone intensity distribution nor in
its autocorrelation, which is equivalent to a HBT type of measurement.
The information gradually appears in the test arm as the degree of
spatial coherence increases. The converse holds for the ghost
diffraction scheme: the information on the phase object can be
retrieved from the correlation between the test and the reference arm
{\em only when the light is spatially incoherent}, and it disappears
when increasing the coherence.  From these results we will conclude
that the claim of Ref.~\cite{abouraddy:2004} is incorrect for the ghost
imaging scheme, which indeed works as a coherent imaging scheme only
{\em because of incoherence}. This claim could be possibly applied
to the HBT scheme, which in the case of a phase object works as a
coherent imaging system only when the light is coherent. 
\par
Thus, there exist remarkable differences between the ghost imaging and
the HBT schemes, which will be clarified for the first time -- to our
knowledge -- in the present work.
\par
The paper is organized as follows. Sec.~\ref{sec:Setup} describes the
experimental setup.  Sec.~\ref{sec:Theoretical-results} is devoted to
theoretical results, with Sec.~\ref{sec:Theory-behind-ghost} giving
the basics of the theory of the ghost diffraction,
Sec~\ref{sec:Basic-theory} discussing the properties of the chosen
object, and Sec.~\ref{sec:Prel-investigation-t} presenting some
numerical results used as a guideline for the experimental
implementation. In fact, the speckles needed are so small that we have
to use a near-field scattering technique to produce them. This
technique, along with its experimental implementation is described in
Sec.~\ref{sec:NFS}.  Sec.~\ref{sec:Exper-results-disc} finally
presents the experimental results of ghost diffraction and of the HBT
scheme with incoherent illumination, while
Sec.~\ref{sec:Exper-results-disc2} present results illustrating the
transition from incoherent to coherent illumination.
Sec.~\ref{sec:Conclusion} concludes.

\section{The setup}
\label{sec:Setup}

\begin{figure}[tb]
    {\scalebox{.55}{\includegraphics*{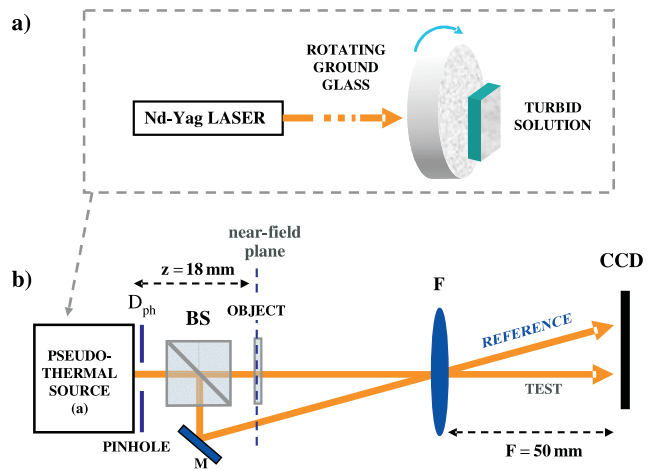}}}
\caption{Experimental setup for ghost diffraction of a pure phase
  object using classical pseudo-thermal light. See text for details.}
\label{fig:setup}
\end{figure}  
The ghost diffraction setup used in our experiment is schematically
shown in Fig.~\ref{fig:setup}.  A pseudo-thermal source generates a
chaotic speckle beam, which enters a balanced 50/50 BS (the side of
the glass cube is 12.5 mm). The pseudo-thermal source
[Fig.~\ref{fig:setup}(a)] consists of a large collimated laser beam
(frequency doubled Nd:YAG laser, $\lambda=0.532~\mu$m,
diameter $D\approx 10$~mm) illuminating a slowly rotating ground
glass followed by a square cell containing a colloidal turbid
solution. The transversal size of the source is delimited by a
pinhole of diameter $D_{\rm ph}=4.5$~mm, which is placed at the
exit face of the turbid cell. A detailed description of the thermal
source together with the features of the speckles that it generates
will be given in Sec.~\ref{sec:NFS}.

The BS divides the pseudo-thermal radiation into two "twin" speckle
beams: the transmitted one is used for the test arm and is sent
onto the object located  right after the BS at a distance $z=18$~mm
from the source. The reflected beam is used for the reference arm;
the mirror $M$ deflects it towards the detector in a direction forming 
a small angle with the test arm. Note that, thanks to the
double reflection of the reference arm, the test and reference
beams on the detector are not mutually reversed, but are each others
replica.

Both beams are collected by the central part of the lens $F$ (focal
length $F=50$~mm) and their intensity distributions are detected by a
charged coupled device (CCD) sensor placed in the focal plane of
the lens. Due to the small angle between the two arms, the
corresponding light spots fall onto different non-overlapping regions
of the CCD sensor.

The data are acquired with an exposure time $\approx 0.1$~ms much
shorter than the speckle coherence time $\tau_{\rm coh} \approx
10$~ms (see Sec.~\ref{sec:NFS}), allowing the recording of high contrast
speckle patterns. The frames are grabbed at a rate of ~5 Hz or
smaller, so that each data acquisition corresponds to uncorrelated
speckle patterns.


\section{Theoretical results}   
\label{sec:Theoretical-results}
\subsection{General theory behind the setup}
\label{sec:Theory-behind-ghost}

In this Section we present the basic theory behind the ghost imaging
setup of Fig.~\ref{fig:setup}.

The general theory of ghost imaging has been explained in detail in
Refs.~\cite{thermal,gatti:2006}. We summarize here the main points:
\begin{enumerate}
\item The collection time of our measuring apparatus is much smaller than
the time $\tau_{\rm coh}$ over which the speckle fluctuate. Hence all
the field operators are taken at equal times, and the time argument is
omitted in the treatment.
\item The speckle beam generated by the pseudo-thermal source is
described by a thermal mixture, characterized by a Gaussian field
statistics. Any correlation function of arbitrary order is thus
expressed via the second-order correlation function
\begin{equation}
\Gamma(\xv,\xp)=
\langle a^\dagger (\xv) a (\xp) \rangle,
\label{gamma}
\end{equation}
where $a$ denotes the boson field operator of the speckle beam.
Notice that we are dealing with classical fields, so that the field
operator $a$ could be replaced by a stochastic c-number field, and the
quantum averages by statistical averages over independent data
acquisitions. However, we prefer to keep a quantum formalism in order
to outline the parallelism with the quantum entangled beams from PDC.
\item Information about the object is extracted by measuring the
spatial correlation function of the intensities $\langle I_1 (\xv_1)
I_2 (\xv_2) \rangle$, where 1 and 2 label the test and the reference
beam, respectively, and $ I_i (\xv_i)$ are the operators associated to
the number of photo counts over the CCD pixel located at $\xv_i$ in
the $i$th beam. All the information about the object is contained in
the correlation function of intensity fluctuations, which is
calculated by subtracting the {\em background} term $\langle I_1
(\xv_1)\rangle \langle I_2 (\xv_2) \rangle \ $:
\begin{equation}
G(\xv_1, \xv_2) = \langle I_1 (\xv_1) I_2 (\xv_2) \rangle - \langle I_1 (\xv_1)\rangle \langle I_2 (\xv_2) \rangle \; .
\label{eq6}
\end{equation}
\par
By using the input-output relations of the BS, and the standard
properties of Gaussian beams, the main result obtained in
Ref.~\cite{thermal} was
\begin{eqnarray}
G(\xv_1, \xv_2) = |rt|^2
\nonumber\\ \times
\left| \int {\rm d} \xp_1
\int {\rm d} \xp_2  h_1^* (\xv_1, \xp_1) h_2 (\xv_2, \xp_2) \Gamma(\xp_1,\xp_2)
\right|^2 \; ,
\label{eq12}
\end{eqnarray}
where $h_1 $, $h_2$ are the impulse response function describing the optical
setups in the two arms, and $r,t$ the reflection and transmission coefficients of the BS. 
\end{enumerate}

Equation~(\ref{eq12}) has to be compared with the analogous result
obtained for PDC beams \cite{gatti:2003}:
\begin{eqnarray}
  G_{\rm pdc}(\xv_1, \xv_2)  = 
\nonumber\\
\left| \int d \xp_1
\int d \xp_2  h_1 (\xv_1, \xp_1) h_2 (\xv_2, \xp_2) \Gamma_{\rm
  pdc}(\xp_1,\xp_2) 
\right|^2,
\label{eq:pdc}
\end{eqnarray}
where $1$ and $2$ label the signal and idler down-converted fields
$a_1$, $a_2$, and
\begin{equation}
\Gamma_{\rm
  pdc}(\xp_1,\xp_2) = \langle a_1 (\xp_1) a_2 (\xp_2) \rangle 
\label{gamma:pdc}
\end{equation}
is the second-order field correlation between the signal and idler
(also called biphoton amplitude).
\par
Thus, ghost imaging with correlated thermal beams presents a deep
analogy with ghost imaging with entangled beams
\cite{thermal,gatti:2006,bache:2005}: they are both coherent imaging
systems, which is crucial for observing interference from a phase
object, and they offer analogous performances provided that the the
beams have similar spatial coherence properties.  They differ in (a)
the presence of $h_1^*$ at the place of $h_1$ (which in our case turns
out to have no implications) and (b) the visibility, defined as
\begin{equation}
{\cal V}= 
{\rm max}\left[ \frac{  G      } 
{\langle I_1I_2  \rangle     }\right]   =
 {\rm max}\left[ \frac{  G    } 
{ \langle I_1 \rangle \langle I_2 \rangle +  G }\right]     \, .  
\label{visibility}
\end{equation} 
In the thermal case $ G(\xv_1, \xv_2) \le \langle I_1(\xv_1) \rangle
\langle I_2( \xv_2) \rangle $ so that the visibility is never above
$\frac{1}{2}$. Conversely, in the PDC case it can be verified that
$G_{\rm pdc}/ \langle I_1 \rangle \langle I_2 \rangle $ scales as $ 1
+ \frac{1}{\langle n \rangle}$, where $\langle n \rangle $ is the mean
photon number per mode (see, e.g., Ref.~\cite{thermal}). Only in the
coincidence-count regime, where $ \langle n \rangle \ll 1 $, the
visibility can be close to unity, while bright entangled beams with
$\langle n \rangle \gg 1$ show a similar visibility as the classical
beams. However, despite never being above $\frac{1}{2}$ in the
classical case, we have shown \cite{thermal,ferri:2004,gatti:2006}
that the visibility is sufficient to efficiently retrieve information.
 
The result of a specific correlation measurement is obtained by
inserting into Eq.~(\ref{eq12}) the propagators describing the setup.
In the case of the ghost diffraction scheme of Fig.~\ref{fig:setup}:
$h_1(\xv_1,\xp_1)=(\im\lambda F)^{-1} e^{-\im \xv_1 \cdot \xp_1 k/ F }
\tobj(\xp_1)$, where $\tobj(\xv)$ is the object transmission function,
$k=2\pi/\lambda$, and $h_2(\xv_2,\xp_2)=(\im\lambda F)^{-1} e^{-\im
  \xv_2 \cdot \xp_2 k/ F } $. We get
\begin{widetext}
\begin{eqnarray}
 G(\xv_1, \xv_2)  \propto
\left| \int {\rm d} \xp_1
\int {\rm d} \xp_2 \,  e^{i \left(\xv_1 \cdot \xp_1 -\xv_2\cdot
    \xp_2\right)k/F}   
T_{\rm obj}^*(\xp_1) \Gamma_{\rm n} (\xp_1,\xp_2) \right|^2 
 \label{diff0}    \\
= 
\left|2\pi 
\int {\rm d} \xiv 
\ttobj \left[ (\xv_1 -\xv_2-\xiv) k/F \right] \Gamma_{\rm f}
(\xv_2, \xv_2 +\xiv )  
\right|^2 ,
\label{diff1} 
\end{eqnarray}
\end{widetext}
where $\Gamma_{\rm n}$ and $ \Gamma_{\rm f}$ denote the second order
field correlation function defined by Eq.~(\ref{gamma}), as measured
at the object near-field and far-field planes, respectively; $\ttobj
(\q) = \int \frac{{\rm d} \xv}{2\pi} e^{-\im \q \cdot \xv} \tobj(\xv)$
is the amplitude of the diffraction pattern from the object, and
Eq.~(\ref{diff1}) is obtained from Eq.~(\ref{diff0}) with some simple
passages.  

We notice first of all that the result of a correlation
measurement is a convolution of the diffraction pattern amplitude with
the far-field coherence function $ \Gamma_{\rm f}$. Hence the {\em
  far-field coherence length} (denoted by $\sff$) determines the
spatial resolution in the ghost diffraction scheme: the smaller the
far-field speckles, the better resolved is the pattern. In the limit
of speckles smaller than the scale of variation of the diffraction
pattern, we can approximate the far field coherence function as
$\Gamma_{\rm f} (\xv_2, \xv_2 +\xiv ) \approx \delta (\xiv) \langle
I_{\rm f} (\xv_2) \rangle$, where $\langle I_{\rm f} (\xv_2) \rangle$ is the
intensity profile of the input speckle beam as observed in the far
field [notice that in our ghost-diffraction setup $\langle I_{\rm f} (\xv_2)
\rangle=\langle I_2 (\xv_2) \rangle$, a part for some trivial
proportionality factor].  In this limit we get
\begin{eqnarray}
G (\xv_1, \xv_2) \propto 
\left|  \ttobj [ (\xv_1-\xv_2) k/F ] 
\right|^2  \,
\langle I_{\rm f} (\xv_2) \rangle^2,
\label{diff2}
\end{eqnarray}
which means that the diffraction pattern of the object can be observed
in the correlation function, when this is evaluated as a function of
$\xv_2$, for fixed $\xv_1$.  The diffraction pattern is modulated by
$\langle I_{\rm f} (\xv_2) \rangle^2$: hence, in order to
obtain the whole diffraction pattern, the far-field intensity
distribution must be sufficiently broad, so that  $\langle I_{\rm f} (\xv_2) \rangle$ is nonzero
in the region where 
the diffraction pattern is nonzero.
  It can be easily seen that
this condition turns out to be equivalent to requiring spatial
incoherence of the speckle beam illuminating the object, that is,
the {\em near-field coherence length} (denoted here as $\snf$) must be
small as compared to the object scale of variation. 
\par
Eq.~(\ref{diff2})
shows that the diffraction pattern can be also obtained by plotting
the correlation as a function of the distance between the two points.
By fixing this distance as $\rv = \xv_1- \xv_2$, and varying the pixel
positions in both arms as $\xv_1$ and $\xv_2= \xv_1- \rv$ we perform a
{\em spatial average} of the correlation function, which amounts to
measuring \cite{bache:2004,bache:2004a}
\begin{equation}
\int {\rm d} \xv G (\xv, \xv -\rv) \propto \left|  \ttobj [ \rv k/F ] 
\right|^2  \int {\rm d} \xv 
\langle I_{\rm f} (\xv-\rv) \rangle^2 \; .
\label{spatialave}
\end{equation}
If the spatial average is performed over large enough regions, the
integral on the right hand side does not depend on $\rv$ and is a
constant.  As already pointed out in Ref.~\cite{bache:2004a}, in this
case there is no need of demodulating the correlation by the mean
intensity in order to obtain the diffraction pattern.

Second of all, and most important for the results presented here:
since the Fourier transform of the amplitude of the object transmission function is involved in
Eq.~(\ref{diff1}), the imaging scheme is coherent despite the fact that
the beams are incoherent. Thus, ghost diffraction of a pure phase
object can be realized with spatially incoherent pseudo-thermal beams.

\par
Quite different are the results for the HBT-type scheme, where the BS is effectively placed after the object.
In this case  the reference kernel changes to
$h_2(\xv_2,\xp_2) = h_1(\xv_2,\xp_2) =(\im\lambda F)^{-1} e^{-\im
  \xv_2 \cdot \xp_2 k/F} \tobj(\xp_2)$, and a result of correlation measurement gives
\begin{widetext}
\begin{eqnarray}
G_{\rm HBT}(\xv_1, \xv_2) \propto
\left| \int {\rm d} \xp_1
\int {\rm d} \xp_2 \,  e^{i k/F \left(\xv_1 \cdot \xp_1 -\xv_2\cdot \xp_2\right)} 
T_{\rm obj}^*(\xp_1) T_{\rm obj} (\xp_2) 
\Gamma_{\rm n} (\xp_1,\xp_2)
\right|^2 \; .
\label{HBT1}
\end{eqnarray}
In the limit of spatially incoherent light $\Gamma_{\rm n}(\xp_1,\xp_2)
\to \delta (\xp_1-\xp_2) \langle I_n(\xp_1) \rangle$, and Eq.~(\ref{HBT1}) can be re-casted as:
\begin{eqnarray}
G_{\rm HBT} (\xv_1, \xv_2) &\propto&
\left|
\int {\rm d} \xiv
 \,
\ttobjHBT \left[ (\xv_2-\xv_1 +\xiv) k/F\right]  \Gamma_{\rm f}
(\xv_2, \xv_2 +\xiv )  
\right|^2 \, .
\label{diffHBT} 
\end{eqnarray}
\end{widetext}
By comparing with the result of Eq.~(\ref{diff1}) for ghost diffraction, we
see that the HBT scheme only gives information about the Fourier
transform of the modulus squared of the object transmission function: in the limit of incoherent light 
the imaging scheme is incoherent. Thus, the phase information about the object is
lost and the HBT scheme is able to see interference only from
absorptive objects.
\par
We can now consider the opposite limit, of spatially coherent light illuminating the object, achieved 
when the coherence length $\snf$ (the speckle size)  at the object plane is large compared to the
object size. In this case,
 the HBT scheme allows to retrieve the diffraction pattern
even of a pure phase object \footnote{ It should be remarked that
 the HBT correlation vanishes in the limit of full 
  coherence, where the second-order correlation $\langle
  I_1 I_2\rangle$ factorizes so that $G_{\rm HBT}=0$. However, this presumes
  both spatial and temporal coherence. What we intend here, instead, is simply that we have 
large speckles on the object, implying thus spatial  coherence;  the light is however temporally
  incoherent.}. In this limit, in fact, the coherence function
$\Gamma_{\rm n}$ can be taken as roughly constant over the regions of
integration in Eq.~(\ref{HBT1}), which hence gives:
\begin{eqnarray}
  \label{eq:cauto}
G_{\rm HBT} (\xv_1, \xv_2) \propto |\ttobj(\xv_1 k/F )|^2 \nonumber \\
  \times
  |\ttobj(\xv_2 k/F )|^2 .
\end{eqnarray}
Evidently, if we fix $\xv_1$ and evaluate the correlation as a
function of $\xv_2$ we observe the diffraction pattern, even of a pure
phase object.

Notably, in the ghost diffraction scheme no diffraction pattern
appears in the correlation as a function of the reference pixel
position $\xv_2$ for spatially coherent light. In the limit of spatial
coherence, Eq.~(\ref{diff0}) factorizes into the product of two
integrals, one showing the diffraction pattern of the object in arm 1,
as a function of $\xv_1$, the other showing the mean far-field
intensity profile in arm 2.  This can be readily seen also from
Eq.~(\ref{diff1}), where the limit of spatially coherent light at the
object plane (limit of a single large speckle illuminating the object)
amounts to $\langle I_{\rm f} (\xv_2) \rangle \to \delta (\xv_2)$.

In practice, the classic HBT scheme uses the cross-correlation of two
beams split on a BS after the object as a convenient way of measuring
the auto-correlation of the beam transmitted through the object, as,
e.g., done in Ref.~\cite{saleh:2000}.  We will actually measure the
auto-correlation of the light in the test arm in the focal plane of
the lens, defined as
\begin{equation}
C_{\rm auto}(\xv,\xv')=\langle I_1(\xv)I_1(\xv')\rangle-\langle
I_1(\xv)\rangle\langle I_1(\xv')\rangle \,. 
\label{eq:cautodef}
\end{equation}
 A part from a small shot-noise contribution
at $\xv'=\xv$ and some irrelevant proportionality factors, the results expected from 
such a measurement
coincide with those of the HBT schemes described by Eq.~({\ref{HBT1}), and in the proper limits by 
Eqs.~(\ref{diffHBT}), (\ref{eq:cauto}). 
\par
This comparison between the HBT and the ghost-diffraction schemes
indicates that the measurement of the autocorrelation function is a
valid test to prove whether or not a given object alters the phase of the
incoming light or not: in the presence of spatially incoherent
light, the ghost diffraction scheme fully preserves the information
about the object phase in the diffraction pattern, whereas this
information is lost in the autocorrelation. For a pure phase object,
no interference pattern at all should appear in the autocorrelation.
However, as it will become clearer in the next sections, the
autocorrelation is extremely sensitive to the degree of spatial
incoherence in the beam: even a small partial coherence in the
incoming beam is sufficient to preserve some phase information in the
autocorrelation function.
\par
We finally stress that despite having discussed the ghost diffraction
results in the framework of classical speckle light [i.e., for which
Eq.~(\ref{eq12}) holds], the results~(\ref{diff1})-(\ref{eq:cauto}) hold
also for the entangled case [for which Eq.~(\ref{gamma:pdc}) holds].

\subsection{The object}
\label{sec:Basic-theory}

We have chosen a commercially available TGBS as our pure phase object.
It is well known that such a device transmits the incoming light (and
has close to zero absorption/reflection) so that in the far zone of
the object several distinct peaks are observed.  This is because the
diffraction angle obeys the thin grating equation
\begin{equation}
  \label{eq:grating-equation}
  \sin(\theta_n)=n\lambda/d, \quad n=0,\pm 1,\pm 2,...
\end{equation}
This equation holds when the incoming light is a plane wave normally
incident on the grating, and tells that the light is observed in
several orders $n$ of the diffraction angle
$\theta_n$, and that the location of the diffraction peaks is found
according to the ratio $\lambda/d$ where $\lambda$ is the wavelength
of the light and $d$ is the period of the grating.
\par
When observing the intensity distribution of the transmitted light in
the far zone, the strength of the $n$th diffraction peak is
$\eta_n=|c_n|^2$, where $c_n$ are the diffraction coefficients.  Since
a grating can be thought of as an infinite repetition of a single
period of the grating $\tobj$, we may use Fourier series theory to
write the diffraction coefficients \cite{goodman:1996}
\begin{equation}
  \label{eq:2}
  c_n=\frac{1}{d}\int_0^d dx \tobj(x)e^{-i n 2\pi x/d}.
\end{equation}
\par
A TGBS is a grating made of a completely transparent material which
has grooves cut on the exit side so that a phase difference is
imposed on the field exiting the grating depending on the position.
This phase difference can be expressed through the groove depth
$\delta$ as $\Delta \phi=(n_g-1)2\pi \delta/\lambda$, where $n_g$ is
the refractive index of the grating material. Thus, $\delta$ can be
chosen to give the desired phase shift.
\par
Typically, square gratings of width $a$ (and period $d$) are used for
TGBS, so that within the period $d$ the object transmission function is
\begin{eqnarray}
  \label{eq:square}
  \tobj(x)=\left\{\begin{matrix}
e^{i\Delta \phi} &  0\leq x \leq a\\
1,  & a< x \leq d
\end{matrix}
\right.~.
\end{eqnarray}
Calculating the diffraction coefficients gives
\begin{eqnarray}
  \label{eq:4}
  \eta_0&=&1+\sin^2(\Delta \phi/2)\frac{4a}{d}(a/d-1)\\
  \eta_n&=&\sin^2(\Delta \phi/2)\frac{4a^2}{d^2}\mathrm{sinc}^2(\pi n
  a/d), \quad n=\pm 1,\pm 2, ... 
\end{eqnarray}
Choosing $\Delta \phi$ and the ratio $a/d$ properly one can engineer
the peaks to have the desired distribution.
\par
Our TGBS is from Edmund Optics and has 80 grooves per mm (stock number
NT46-069), i.e., $d=12.5~\mu $m. It is designed for $\lambda=633$~nm to
have 25 \% of the power in the 0 and $\pm 1$ peaks, and 5 \% in the
$\pm 2$ order peaks. This means $\eta_0=\eta_1$ and
$\eta_1/\eta_2=5$, implying that $\Delta \phi=0.71\pi$ and that
$a/d=\arccos(1/\sqrt{5})/\pi$ giving $a=d/2.84=4.4~\mu$m. The
smallness of $a$ sets a limit for the relative coherence of the object
beam, as discussed in detail in what follows.
\par
Since we are using a frequency doubled Nd:YAG laser with
$\lambda=532$~nm the phase difference is instead $\Delta
\phi=\frac{633}{532}0.71 \pi\simeq 0.84 \pi$. This implies
$\eta_1/\eta_0=2.2$, i.e., the central peak should be roughly a factor
of 2 weaker than the $\pm 1$ order peaks when using light at this
wavelength.

\subsection{Preliminary investigation through numerics}
\label{sec:Prel-investigation-t}

\begin{figure}  [tb] 
{\scalebox{.7}{\includegraphics*{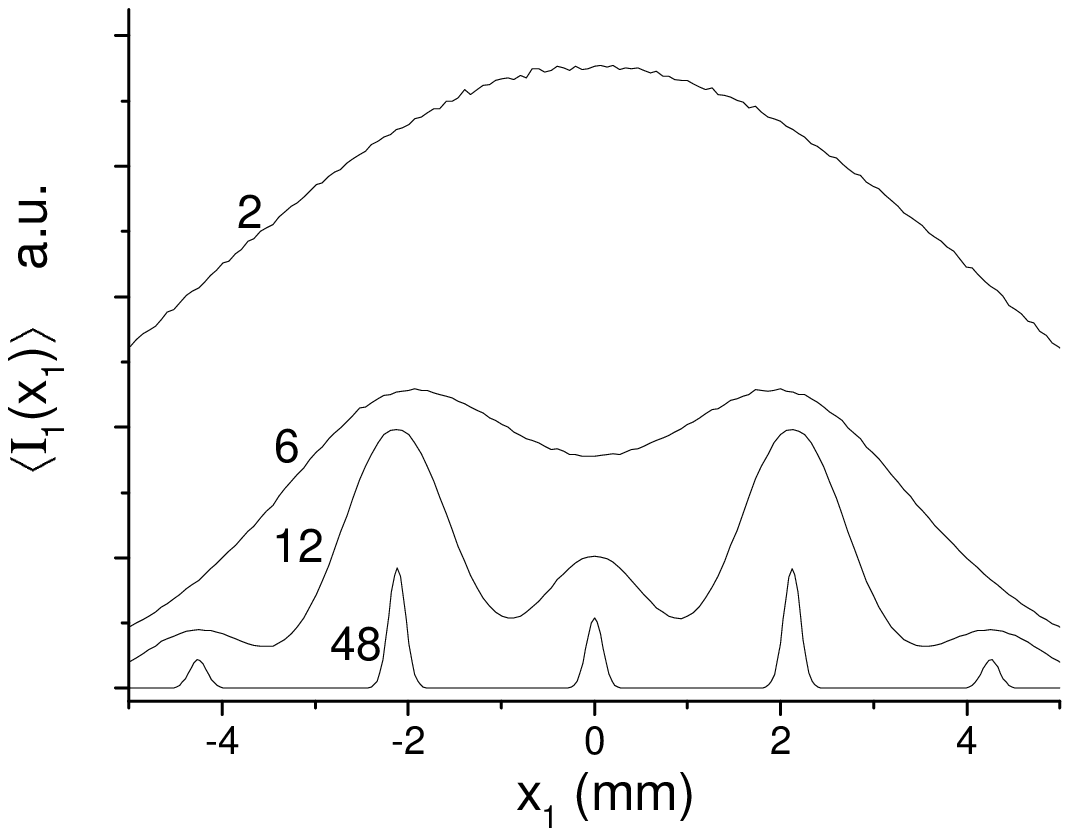}}}
\caption{Numerical simulations demonstrating the transition between
  coherent and incoherent illumination of the object.  Average
  intensity $\langle I_1\rangle$ in the far field of the test arm for
  different values of the speckle size at the object plane
  $\snf=2,6,12$, and $48~\mu$m Numerical parameters: we use $1024$
  grid points, and 32 pixels per groove period; $d=12.5~\mu$m;
  $a=4.2~\mu$m, $\Delta \phi=0.84\pi$; $\sff=80~\mu$m; averages are
  done over $10^5$ independent realizations.}
\label{fig:coherence}
\end{figure}  

In this section we will discuss some numerical results we used to
select the size of the speckles we would need in the object plane in
order to render the beam spatially incoherent. The numerical
simulations were done using the method described in Ref.~\cite{gatti:2006}.
Essentially, we Fourier transform noise convoluted with a Gaussian to
obtain a certain speckle size in the object plane $\snf$ (the size of
which is controlled by the waist of the Gaussian).  Imposing a
pinhole of size $D_{\rm ph}$ on this field and performing another
Fourier transform gives a speckle field as observed in the far field
of the object plane; the speckle size there $\sff$ is determined by
$D_{\rm ph}$ and is therefore controlled independently of $\snf$.  The
object used in the simulations was a purely transmissive square
grating chosen to mimic the object predicted theoretically in
Sec.~\ref{sec:Basic-theory}, see Fig.~\ref{fig:coherence} for specific
parameters.
Since ghost imaging implies that coherent imaging is done with
incoherent light, it is important that the light is really
incoherent relative to the object
details. Thus, in a direct observation of the test arm far field
average intensity, $\langle I_1\rangle$, we should not be able to see
the diffraction pattern of the object. As shown in
Fig.~\ref{fig:coherence} when $\snf \approx 2 \mu$m $\langle I_1\rangle$
does not reveal any information about the diffraction pattern. Thus,
such a speckle size corresponds to practically incoherent illumination
of the object. However, as the speckle size is increased
($\snf=12~\mu$m) $\langle I_1\rangle$ reveals more and more
information about the diffraction pattern, corresponding to partially
coherent illumination. For large speckle sizes ($\snf=48~\mu$m)
$\langle I_1\rangle$ is very close to the analytical diffraction
pattern and the illumination is close to being completely coherent
\footnote{In fact, the observed diffraction pattern for a given
  speckle size can be understood as the Fourier transform of not the
  infinitely extended grating, but instead the part of the grating
  that the speckle is able to ``see'' due to its finite coherence.
  Thus, e.g., $\langle I_1\rangle$ obtained for $\snf=48~\mu$m
  corresponds (roughly) to the diffraction pattern of a transmission
  grating with 4 periods.}.

\par
\begin{figure*}[htb]
{\scalebox{.75}{\includegraphics*{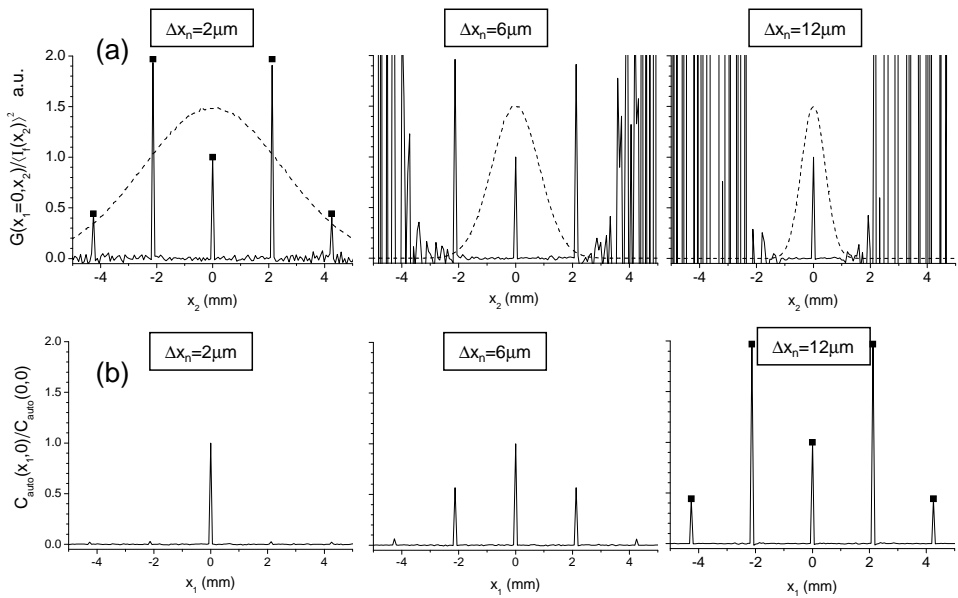}}}
\caption{Numerical simulations demonstrating the transition between
  coherent and incoherent illumination of the object.  (a) The
  cross-correlation (solid line) normalized by the square of the mean
  intensity in the reference arm (the latter is shown by the dotted
  line) as obtained for different sizes $\snf$ of the near field
  speckles. (b) The corresponding autocorrelation functions of
  the test arm intensity.  The black squares indicate the peak values
  of the analytical diffraction pattern. Parameters: as in
  Fig.~\ref{fig:coherence}.}
\label{fig:coherence2}
\end{figure*}  
As shown in Sec.~\ref{sec:Theory-behind-ghost}, if the test beam is
incoherent with respect to the object, the autocorrelation of $I_1$
should not reveal at all the diffraction pattern of a pure phase
object, while the pattern should appear in the cross-correlation.
Fig.~\ref{fig:coherence2} compares the cross-correlation function
$G(\xv_1,\x_2)$
and the autocorrelation $C_{\rm auto}(\xv_1,\xp_1)$ of the test arm
intensity, for different sizes of the near-field speckles. Both
functions are calculated for a fixed $\xv_1$. For $\snf=2.0~\mu$m the
cross-correlation shows a diffraction pattern that, as we verified, is
very close to the pattern analytically calculated. On the contrary,
the autocorrelation shows very little information in the sidebands;
the $n=1$ sideband is 2.4\% of the central ($n=0$) peak. This confirms
that this speckle size corresponds to practically incoherent
illumination. The fact that some information is observed in the
sidebands at all is because the 2.0 $\mu$m speckle on the scale of the
object is not vanishingly small, but merely small.  By increasing the
coherence of the light, we notice that the object information
disappears from the cross-correlation and appears in the
autocorrelation, in agreement with Eq.~(\ref{eq:cauto}) and with the
discussion in Sec.~\ref{sec:Theory-behind-ghost}.  For $\snf=12~\mu$m
the sidebands present in the autocorrelation coincide almost
completely with the analytical diffraction pattern, showing thus that
the autocorrelation function is sensitive to the presence of even a
small partial coherence of the light.
\par
To conclude this discussion, we have to choose a speckle size at the
object plane around   $2~\mu$m in order to have beams that are
truly incoherent with respect to the object details, so that
information about the object is revealed neither in the
autocorrelation nor the average of the test beam far-field intensity.

\section{Experimental results}
\subsection{Pseudo-thermal source and speckles from near-field scattering}
\label{sec:NFS}

As shown in the previous section, the physical size of the TGBS
requires a speckle size at the object plane $\snf$ less than the
finest object detail, i.e., less than 4.4 $\mu$m. Such a small size is
not easy to achieve, but we managed to create speckles with
$\snf=2.0~\mu$m by placing the object very close to our pseudo-thermal
source ($z=18$~mm, practically limited only by the physical size of
the half-inch cube BS). The speckles generated this way are the
so-called {\em near-field speckles} (NFS) \cite{giglio:2000,giglio:2004},
which are remarkably different from the classical far-field speckles (FFS), 
whose size is determined by the well-known Van Cittert-Zernike theorem 
\cite{goodman:1975}.

This part of the setup is quite different from what we used in our
previous experiments in Refs.~\cite{ferri:2004,gatti:2006,bache:2005}
where the object was in the far zone of the source. Instead, in the
current setup the source is very close to the object, and, as
explained in detail below, at a given point of the object plane the
waves interfering do not originate from the entire illuminated spot.

To understand how NFS are formed, let us first describe the
pseudo-thermal source in some detail.

Our thermal source consists of a laser beam illuminating a slowly
rotating ground glass, followed by a square cell $5$~mm thick,
containing a concentrated solution of latex particles with
  average diameter $\rho=3~ \mu$m. The cell is almost in physical
contact with the rotating glass and on its exit face there is a
{pinhole} (diameter $D_{\rm ph}=4.5$~mm) which determines the
transverse size of the source. The combination of the ground glass and
the turbid solution is an easy and convenient way to generate truly
random speckles. Indeed, the ground glass alone would produce only
partially stochastic speckles because the pattern would be reproduced
after one {turn of the glass}.  On the other hand, the turbid cell
guarantees stochasticity, but if used alone would exhibit a residual
transmitted coherent component which is clearly undesired.

Due mainly to Brownian particle motion (and secondarily to glass
rotation), the speckle pattern generated by the source fluctuates
randomly with time and is characterized by a {coherence} time
$\tau_{\rm coh}$ which can be tuned by varying the turbidity of the
solution. In our case we had $\tau_{\rm coh}\approx 10$~ms.

Multiple scattering occurs inside
the cell, so that the light beam exiting the source 
has a divergence angle $\Theta_{\rm eff}$ larger than 
the angle that would be expected from single scattering. The latter 
is associated to particle diffraction and given by
$\Theta_{\rm dif}\approx{\lambda/\rho}$. The effective value of 
$\Theta_{\rm eff}$ depends on the detailed 
features of the scattering cell, as particle concentration and cell length; 
in practice, we can claim that our thick thermal source
behaves as an ideal thin thermal source characterized by spatial
inhomogeneities or "scatterers" of effective diameter $\rho_{\rm
eff}\approx\rho~(\Theta_{\rm dif}/\Theta_{\rm eff})< \rho$.

When the light generated by such a source is observed at a plane
located at a sufficiently large distance $z$, each point of this plane
is reached by the contributions emerging from the entire radiating
region $D_{\rm ph}$. Under this condition, the stochastic interference
between the (many) different waves gives rise to a speckle-like
intensity distribution ({\em far-field speckle}, FFS), whose
correlation function is described by the Van Cittert-Zernike theorem,
and is characterized by the average speckle size \cite{goodman:1975}
\begin{equation}
 \label{eq:FFS1}
 \Delta x \approx z\frac{\lambda}{D_{\rm ph}}      \qquad{\rm (FFS)}.
\end{equation}
Thus, the requirement for obtaining FFS is $\Theta_{\rm eff} z \gg D_{\rm ph}$, i.e.
\begin{equation}
  \label{eq:FFS2}
  z \gg \frac{D_{\rm ph}~\rho_{\rm eff}}{\lambda}   \qquad{\rm (FFS)}.
\end{equation}
When this condition is not fulfilled, the waves interfering at each
point of the observation plane originate from a region  $D^* \approx
\Theta_{\rm eff}~z \approx \frac{\lambda}{\rho_{\rm eff}}~z$ smaller
than the radiating region $D_{\rm ph}$. Provided that $D^*$ is not
too small, at a given point in the observation plane one would still
get contributions from many different scatterers, which is
sufficient to produce {\em near-field speckles}. Applying again
the Van-Cittert Zernike  theorem with $D^*$  instead of $D_{\rm
ph}$, one gets the remarkable result \cite{giglio:2000}
\begin{equation}
  \label{eq:NFS1}
  \Delta x \approx \rho_{\rm eff} \quad {\rm (NFS)}\, ,
\end{equation}
according to which the average NFS size is only determined by the
effective size of the scatterers, and is independent of both $\lambda$
and $z$.  To fulfill the criteria for NFS we must have (i) $D^* < D$
and (ii) many scatterers inside $ D^*$, e.g., $(D^*/\rho_{\rm
  eff})^2\gg1$. This implies that the distance $z$ must fulfill the
two conditions
\begin{equation}
  \label{eq:NFS2}
z \ll  \frac{D_{\rm ph}~\rho_{\rm eff}}{\lambda}, \quad z\gg
\frac{\rho_{\rm eff}^2}{\lambda} \quad {\rm (NFS)}.
\end{equation}
In our experiment we have $\lambda=0.532~\mu$m, $D_{\rm ph}=4.5$~mm
and $\rho_{\rm eff}\le 3~\mu$m. Thus the second criterium is easily
fulfilled because $\frac{\rho_{\rm eff}^2}{\lambda}<
\frac{\rho^2}{\lambda}=17~\mu$m, so that $z \gg 17~\mu$m is enough.
The first criterium is somehow more difficult to evaluate, because it
requires {more detailed} knowledge of $\rho_{\rm eff}$. However,
our final purpose was to make speckles as small as possible and,
therefore, we set the distance between the object and the
pseudo-thermal source to its minimum value $z=18 $~mm \footnote{Note
  that $z=18$~mm is only the geometrical distance, and the presence of
  the half-inch glass cube BS should be taken into account.  In
  particular, when calculating the size of FFS or of $D^*$, the
  effective distance $z_{\rm eff}$ that one should use is shorter than
  $z$, because of the smaller divergence of light in glass than in
  vacuum. In our case {we estimate} $z_{\rm eff} \approx 14$~mm.},
{which to a large extent meets the first criterium in
  Eq.~(\ref{eq:NFS2})}.

Then we measured the speckle pattern in the
object plane by removing the lens $F$ and inserting a $\times 20$
objective to image the object plane on the CCD (the magnification is
needed because the CCD pixel size is 6.7 $\mu$m). The speckle size
was finally estimated by performing the autocorrelation of such
pattern as shown in Fig.~\ref{fig:nearspeckles}.
\begin{figure}[t] 
   {\scalebox{.6}{\includegraphics*{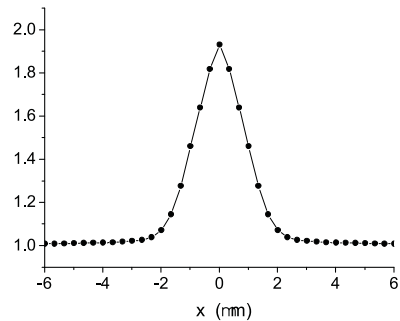}}}
\caption{Spatial autocorrelation function of the speckle beam just
  before the object $\langle I_{\rm n} ( \xp ) I_{\rm n} ( \xp + \x)
  \rangle$. The peak above the baseline has a FWHM of $1.98\pm
  0.02~\mu$m, which gives the characteristic size of the near field
  speckles.}
\label{fig:nearspeckles}
\end{figure}  
The peak above the baseline had a full-width half maximum (FWHM) of
$1.98\pm 0.02~\mu$m.  This gives an estimate of the near-field speckle
size (see Refs.~\cite{ferri:2004,gatti:2006} for more details) $\snf
\approx 1.98~\mu$m, which, as shown in
Sec.~\ref{sec:Prel-investigation-t}, should be small enough to render
the beams incoherent with respect to the object details.
\par
For completeness, we also measured the size of the speckles in the
far-field (the size of the speckles on the CCD) by removing the
objective and reinserting the lens $F$. The procedure gave
$\sff=11.1\pm 0.1~\mu$m. This value actually overestimates the real size of
the far-field speckles, because the CCD pixel size ($6.7~\mu$m)
is too large with respect to the speckles, so that the speckle pattern
undergoes a substantial smoothing.  In any case, the measured $\sff$
determines the spatial resolution of the ghost diffraction pattern,
and in our case turns out to determine the width of the ghost
diffraction peaks.

\subsection{Ghost diffraction versus HBT scheme: case of incoherent
  illumination}
\label{sec:Exper-results-disc}

We performed a first set of measurements keeping the object plane at a
distance $z=18$~mm, thus having speckles at the object plane of size
$\snf \approx 2~\mu$m.
\par
In order to characterize the diffraction pattern created by the TGBS
and to provide a reference for the ghost diffraction pattern, we
performed preliminary measurements with coherent laser light: we
removed the scattering media from the setup of Fig.~\ref{fig:setup},
and recorded the transmitted light of the TGBS in the focal plane of
the lens $F$. This measurement was already not straightforward because
of the large values of the scattering angles of the grating
equation~(\ref{eq:grating-equation}): the $n$th order peaks at angles
$\theta_{n}=n \lambda/d$, are displaced in the far field at positions
$x_n=\theta_n F =nF\lambda/d$.  Using the numbers in our setup we have
$x_{\pm 1}=\pm 2.13$~mm and $x_{\pm 2}=\pm 4.26$~mm, so that the
distance between the two 2nd order peaks is larger than the extension
of our CCD.
We therefore had to do 3 measurements in order to observe all the
peaks: first $n=-2,- 1$, and 0 were observed, then the CCD was shifted
to observe $n=-1,0$, and $+1$ and finally $n=0,+1$, and $+2$. A second
problem was represented by the small width of the diffraction peaks
(few pixels on the CCD) which provided a too poor sampling of the
diffraction pattern. In order to evaluate the relative heights of the
peaks we performed an integration in the region around each peak,
which gave $\eta_1/\eta_0=4$ and $\eta_1/\eta_2=2 $. 
Moreover the diffraction pattern is not symmetric and
for example $\eta_{-1} \ne \eta_1$.  This is somewhat different from
what we expected from the theory, and probably depends on some defects
of fabrication of the TGBS, but it will serve as our reference.
Experimentally we found the peaks to be located at $x_{\pm 1}=\pm
2.22$~mm and $x_{\pm 2}=\pm 4.43$~mm, in good agreement with the
theory.
\par
We also used the coherent illumination to set the origin of our
coordinate systems: in the test arm $\x_1=0$ corresponds to the the
$n=0$ diffraction peak, while in the reference arm the $\x_2=0$ point
is the location of the reference beam. {If the test-arm pixel is
  for example fixed at $\x_1=0$ in the subsequent correlation
  measurements}, we expect that the ghost diffraction pattern will
emerge in a region of the reference arm centered around $\x_2=0$,
while by shifting $\x_1$ the pattern will shift accordingly.
\begin{figure}[t] 
{\scalebox{.45}{\includegraphics*{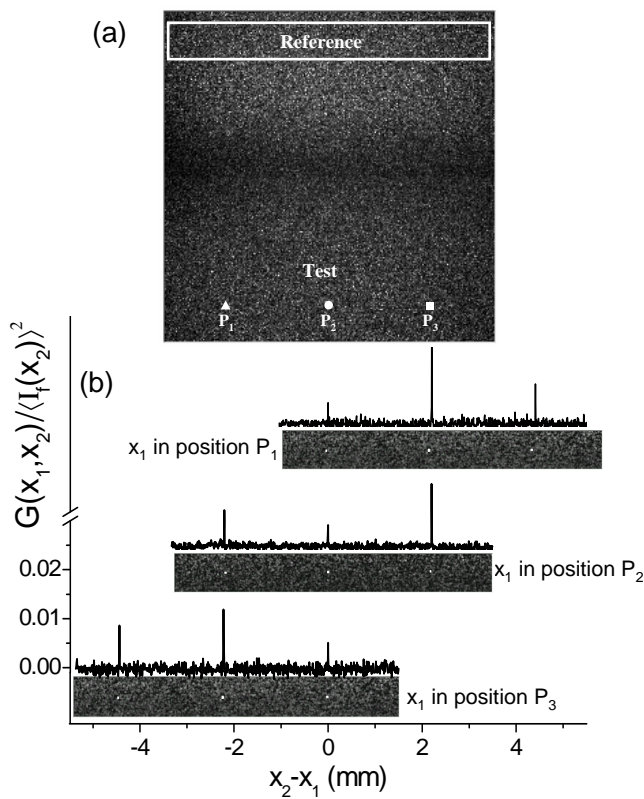}}}
\caption{Experimental demonstration of ghost diffraction of a pure
  phase object using incoherent classical light. (a) Snapshot of the
  speckles recorded by the CCD in the far-field plane.  The reference
  beam is in the upper region and the white frame shows the region
  used for the correlation. The test arm is in the lower region, and
  the white symbols indicate the 3 different single-pixel positions
  used for the correlation. (b) Ghost diffraction patterns
  reconstructed via the cross-correlation between the test and the
  reference arm, measured by locating $\x_1$ at each of the 3 pixel
  positions, and by varying $\x_2$ ($18000$ averages).  The 2-D plots
  of the cross correlation $G(\x_1, \x_2)/ \langle I_{\rm f} (\x_2)
  \rangle^2$ are shown as functions of $\x_2-\x_1$, together with
  their 1-D cut in the horizontal direction. }
\label{fig:cross_single_cut}
\end{figure}  
\begin{figure}[h] 
   {\scalebox{.45}{\includegraphics*{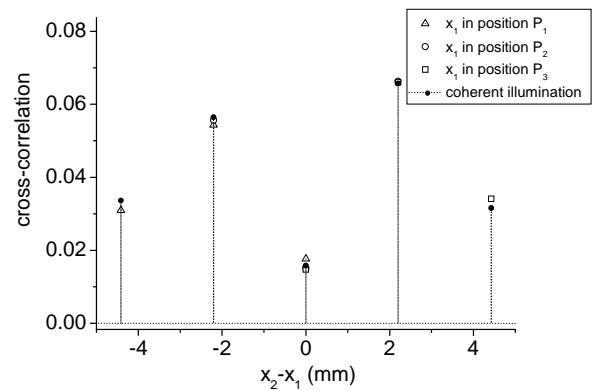}}}
\caption{Quantitative comparison between ghost diffraction and laser
  illumination. The height of the diffraction peaks of the cross
  correlation shown in Fig.~\ref{fig:cross_single_cut} are evaluated
  by performing an integral over the peaks. They are compared with
  peak values measured via coherent laser illumination.}
\label{fig:cross_single_integral}
\end{figure}  

We then reinserted the scattering media to measure the ghost
diffraction pattern of the TGBS from the cross-correlation between the
two arms. A typical snapshot of what we observe on the CCD is shown in
Fig.~\ref{fig:cross_single_cut}(a). The upper part contains the reference
arm intensity. For the correlation, we selected a narrow strip (128
pixel wide) centered around $y=0$ and extending over the entire $x$
axis (1024 pixels). In the test arm no spatial information was extracted since there we collected the light 
from a single fixed pixel. 
Initially we located the pixel in the test arm at the point $\x_1=0$
[pixel at position $P_2$ in Fig.~\ref{fig:cross_single_cut}(a)], and
we measured the cross correlation $\langle I_1 (\x_1=P_2)
I_2(\x_2)\rangle$ as a function of $\x_2$ varying in the region shown
by the white frame in Fig.~\ref{fig:cross_single_cut}(a), by averaging
over $18000$ snapshots. As seen from Eq.~(\ref{diff2}) this gives a
diffraction pattern that is centered on $\x_2=0$. In this case we are
able to reconstruct only the $n=-1,0$ and $+1$ peaks, because the
higher order peaks are outside of the reference region imaged on the
CCD.  In order to reconstruct also the $n=\pm 2$ peaks, we repeated
twice the measurement by shifting the test arm pixel at the positions
$P_1$ and $P_3$, respectively, [see
Fig.~\ref{fig:cross_single_cut}(a)], so that the the diffraction
pattern emerging from the correlation shifts accordingly, as dictated
by Eq.~(\ref{diff2}).

The result of these measurements are shown in
Fig.~\ref{fig:cross_single_cut}(b), which plots the cross correlation
scaled to $\langle I_{\rm f}(\x_2) \rangle^2$.  The 2D reconstructed
diffraction patterns are shown close to their cut along the
x-direction, for each positioning of $\x_1$.  Since each diffraction
peak covers only few pixels, the x-cuts of
Fig.~\ref{fig:cross_single_cut}(b) only give a qualitative image of
the diffraction pattern, but do not allow a quantitative estimation of
the peak heights, due to the poor sampling.  In order to extract the
relative height of the peaks, we located groups of pixels having a
substantial value above the noise floor and added together their
values, which effectively corresponds to integrating over each peak.
The results are shown in Fig.~\ref{fig:cross_single_integral} and
compared with those obtained from coherent laser illumination with the
same technique. We observe that they agree extremely well: we have
successfully created the correct diffraction pattern of the pure phase
object from the correlation.  Also notice that the $n=\pm 1$ peaks are
recorded twice, and the $n=0$ is recorded three times: these overlaps
happen when displacing the single pixel position, and they agree very
well with each other.  Finally, note that the position of the
diffraction peaks agree well with the theoretical prediction in
Sec.~\ref{sec:Basic-theory}.

As discussed in Sec.~\ref{sec:Theory-behind-ghost}, the spatial
average technique provides a a faster and more efficient way of
measuring the ghost diffraction pattern.  In this case the
cross-correlation is measured by varying both $\x_2$ and $\x_1$ for a
fixed $\x_2 -\x_1$. The results of this kind of measurements are shown
in Fig.~\ref{fig:average_cross}, where the upper part of the figure is
a 2-D plot of the measured cross-correlation, and the lower part
displays the integral over the diffraction peaks compared to the laser
illumination results. Also in this case the agreement is excellent.
\begin{figure}[t] 
{\scalebox{.55}{\includegraphics*{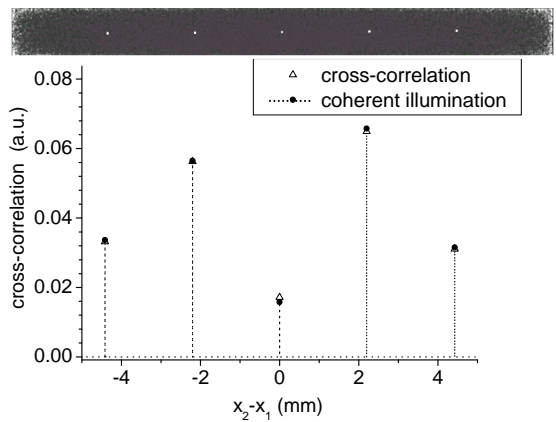}}} 
\caption{Cross-correlation function $\langle I_2(\x_2) I_1 (\x_1)
  \rangle$, calculated with the spatial average technique.  The upper
  part is a 2D plot of the {correlation} function (100 averages). The
  lower plot displays the integrals over the peaks compared with the
  results of laser illumination.}
\label{fig:average_cross}
\end{figure}  
From this figure we notice that only few  averages over
snapshots are needed to reconstruct the diffraction pattern, because a
large number of averages over spatial points is performed, thus
increasing the convergence rate \cite{bache:2004,bache:2004a}.
Moreover the whole diffraction pattern is reconstructed in a single
measurement. Despite being much more efficient than the single-pixel
reconstruction, the spatial average technique does not follow the ghost
imaging original spirit, which assumes that the imaging information is
extracted by only operating on the reference arm. In this case,
spatial information is also extracted from the test beam 1, by varying
the pixel $\x_1$.
  
As a straightforward demonstration of the degree of incoherence of the
beams used, we present in Fig.~\ref{fig:autocorrelation} a measure of
the autocorrelation function in the test arm $C_{\rm auto} (\x_1,
\x_1')$. This is measured by fixing $\x_1'$ at position $P_2$ and by
varying $\x_1$, in the same way as described in
Sec.~\ref{sec:Prel-investigation-t}.
\begin{figure}[h] 
\begin{center}
    {\scalebox{.7}{\includegraphics*{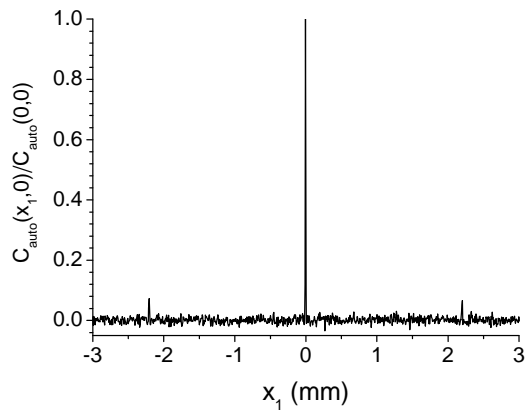}}}
\caption{Autocorrelation function $\langle I_1 (\x_1) I_1 (\xp_1)
  \rangle$ of the test arm, measured by fixing $\xp_1$ at position
  $P_2$ and varying $\x_1$ (10000 averages).}
\label{fig:autocorrelation}
\end{center}
\end{figure}  
Evidently it does not reveal any significant information about the
diffraction pattern. In fact, the first order peaks are barely visible
and are at a level of 8\% of the main peak, {in trend with the
  prediction of} the numerical results of
Sec.~\ref{sec:Prel-investigation-t}. As argued in
Sec.~\ref{sec:Theory-behind-ghost}, this type of measurement is
equivalent to a HBT-type scheme, and it works as an incoherent imaging
scheme when using incoherent light; in this case it is expected to
give no information about a pure phase object.  We can thus conclude
that i) the TGBS is truly a pure phase object and ii) the speckle
light we use is truly incoherent relative to the object.

\subsection{Ghost diffraction versus HBT schemes: case of partially coherent illumination}
\label{sec:Exper-results-disc2}
In this section we present results obtained by gradually increasing
the spatial coherence of the light illuminating the object. This is
achieved by increasing the distance $z$ between the pseudo-thermal
source and the object plane (see Fig.~\ref{fig:setup}).

We performed a second set of measurements with this distance set as
$z=115$~mm. The measured autocorrelation of the light illuminating the
object gave a speckle size $\snf= 14~\mu$m (FWHM of the
autocorrelation peak). The main results obtained in these conditions
are displayed in Fig.~\ref{fig:z115}. In a third set of measurements
the object-source distance was $z= 300$~mm, and the measured speckle
size was $\snf= 33~\mu$m.  Figure~\ref{fig:z300} displays the results in
this case.
\begin{figure}  
\begin{center}
{\scalebox{.6}{\includegraphics*{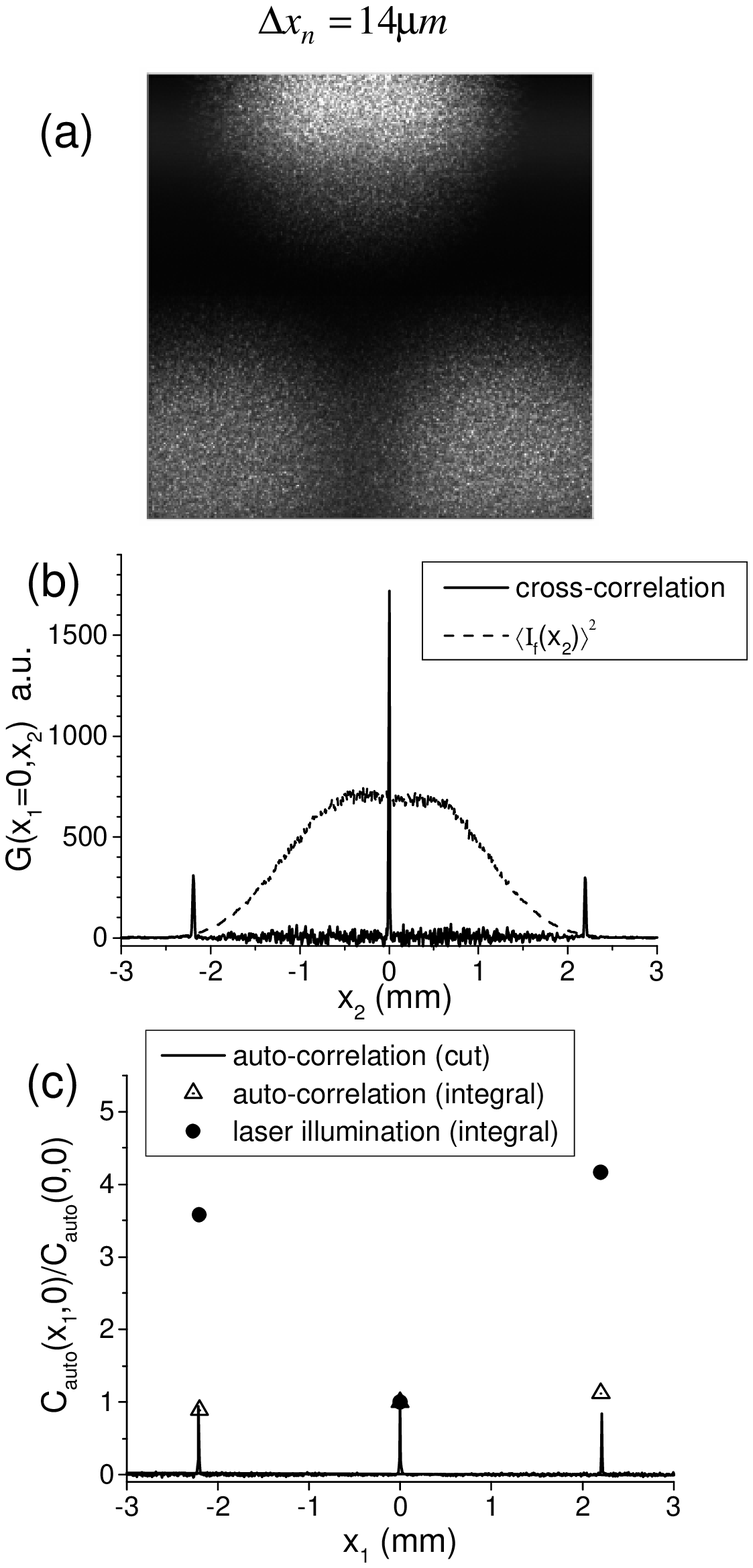}}}
\caption{Case of partially coherent illumination. The distance
  source-object is $z=115$~mm, with $\snf=14~\mu$m. (a) Far-field
  speckle distribution in a single snapshot (upper part: reference
  arm, lower part:test arm).  (b) Cross correlation of the test and
  reference arm, for $\x_1$ fixed at the origin and 30000 averages.
  The dashed curve shows the reference intensity squared.  (c)
  Autocorrelation of the test arm light intensity. Full line:
  horizontal section of the autocorrelation function.  Open triangles:
  peak values of the autocorrelation (integral over the peaks).
  Circles: peak values measured with coherent laser illumination.  }
\label{fig:z115}
\end{center}
\end{figure}  
\begin{figure}  
\begin{center}
{\scalebox{.6}{\includegraphics*{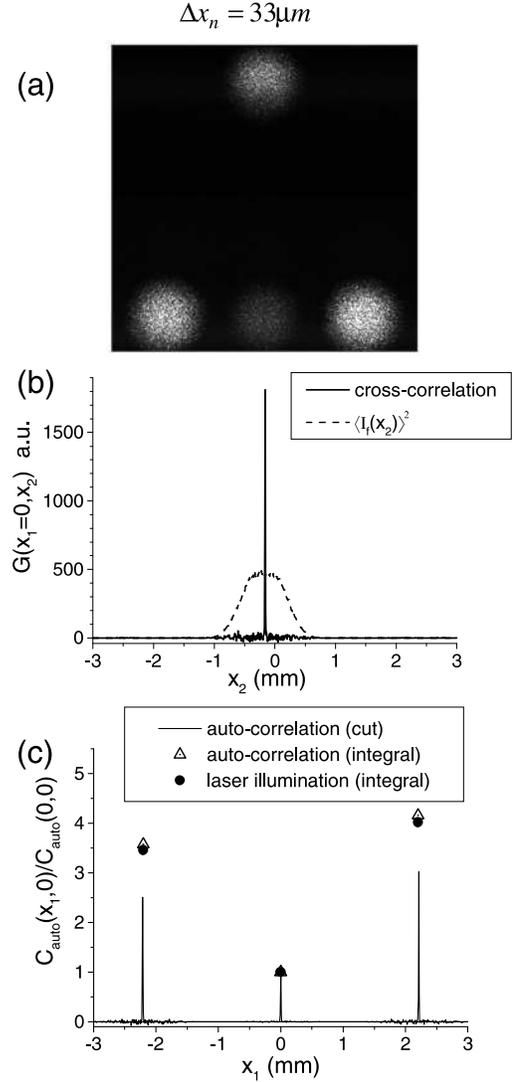}}}
\caption{The spatial coherence of the light illuminating the object is
  further increased: $z=300$~mm, $\snf=33~\mu$m. Frames (a)-(c)
  display the same quantities as in Fig.~\ref{fig:z115}.}
\label{fig:z300}
\end{center}
\end{figure}  

By increasing the source-object distance the light gains some partial
coherence relative to the object. {This is already evident} by the
distribution of the speckles recorded in the far field of the test
arm, shown in the lower parts of frames (a) in Figs.~\ref{fig:z115} and
\ref{fig:z300}. Differently from the case of incoherent illumination,
where the mean intensity distribution of the test arm is almost flat
[see Fig.~\ref{fig:cross_single_cut}(a)], two broad peaks in
correspondence of the $n=\pm 1$ diffraction orders are now clearly
distinguishable in the speckle distribution. As the coherence of the
light increases [Fig.~\ref{fig:z300}(a)] they become narrower and more
pronounced. Notice that the 0 order peak is barely visible in these
plots because its intensity is lower {(as dictated by the TGBS)}.

The cross-correlation between the test and the reference arm, obtained
by fixing $\x_1$ at position $P_2$, is plotted in frames (b) of
Figs.~\ref{fig:z115} and \ref{fig:z300} . We see that by increasing
the coherence of the light, the height of the $\pm 1$ diffraction
peaks decreases with respect the 0 order peak, and the diffraction
pattern gradually disappears from the cross-correlation. Notice
that these plots shows the "bare" cross-correlation function
$G(\x_1,\x_2)$ [i.e., there is no scaling factor $\langle I_{\rm f}
(\x_2) \rangle^2$]. As predicted by Eq.~(\ref{diff2}), the correlation
scales with the square of the mean intensity of the reference arm,
whose profile is plotted by the dashed lines in the figures. By
increasing the near-field coherence, the far-field intensity spot
becomes narrower, until the mean intensity vanishes in the region
where the higher order diffraction peaks should emerge.  In principle,
the correct height of the diffraction peaks could be recovered by
dividing the correlation by $\langle I_{\rm f} (\x_2) \rangle^2$, but
this operation also amplifies the noise in the regions where the
intensity level is low.
  This is evident when the cross-correlation is normalized, as
  shown in Fig.~\ref{fig:crossnorm} (see also Fig.~\ref{fig:coherence2}), and we notice that in the case
of $\snf=14~\mu$m the $\pm 1$ peaks can be almost reconstructed, while
for $\snf=33~\mu$m they disappear in the noise.
\begin{figure}  
{\scalebox{.5}{\includegraphics*{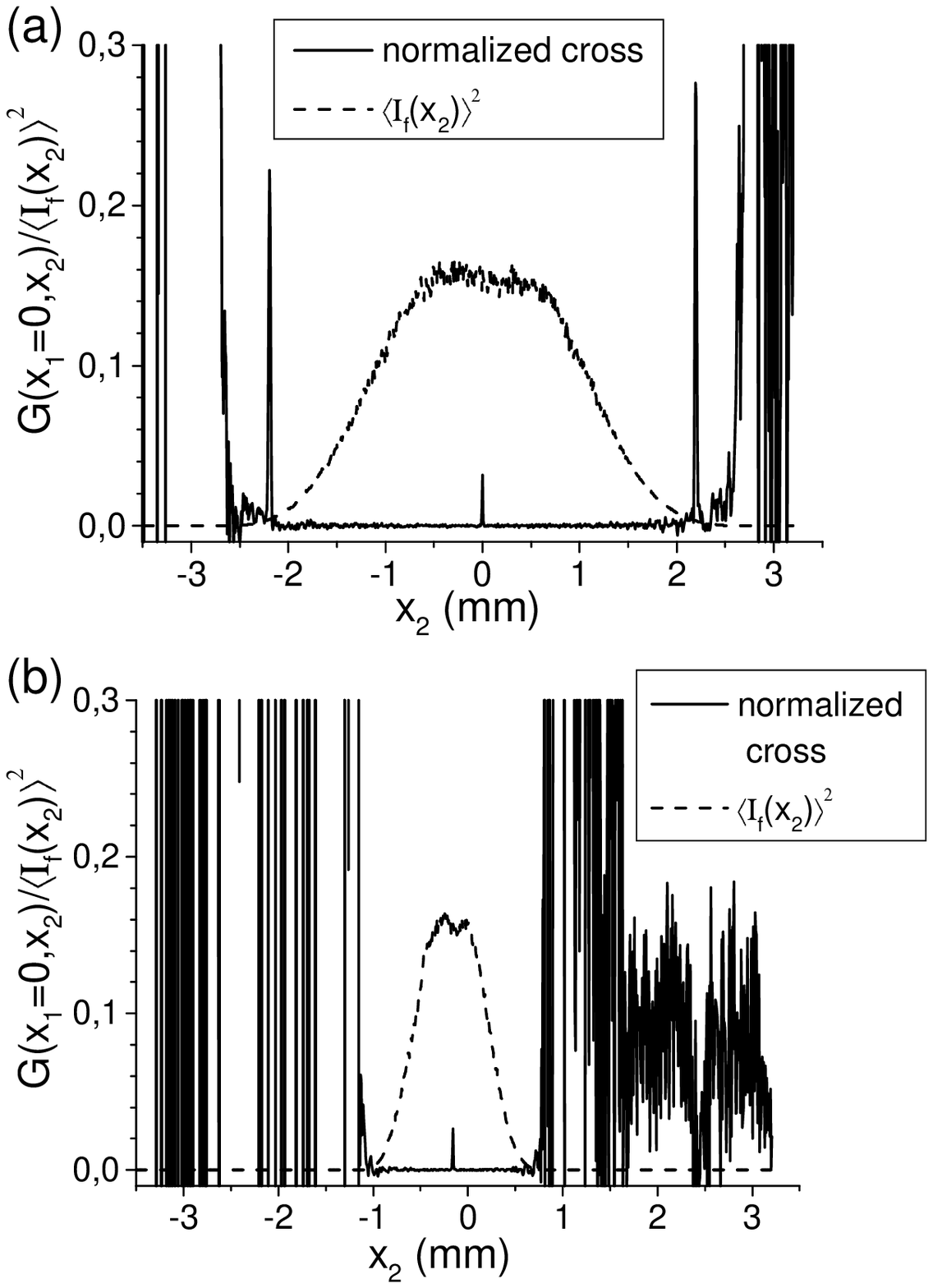}}}
\caption{Cross-correlation function normalized to the square of the
  mean intensity in the reference arm.  (a) $z=115$~mm, $\snf=14~\mu$m:
  the $\pm 1$ peaks can still be reconstructed. (b) $z=300$~mm,
  $\snf=33~\mu$m: the diffraction peaks disappear in the noise. 30000
  averages are performed for both frames.}
\label{fig:crossnorm}
\end{figure}  

In other words, by increasing the coherence, the signal-to-noise ratio
for the reconstruction of the higher order peaks decreases, and the
information about the diffraction pattern becomes less and less
accessible.  To make this argument more formal, we remind that the
signal-to-noise ratio is proportional to the visibility defined by Eq.
(\ref{visibility}), as derived in Ref.~\cite{gatti:2006}.  By using
Eq.~(\ref{diff1}), we can readily conclude that {for $G$ small} the
visibility of the $n$th order peak, located at $\x_2= \bar{\x}_2 $, is
${\cal V} \approx G(\x_1,\bar{\x}_2 )/[ \langle I_1(\x_1) \rangle \langle I_2
(\bar{\x}_2 ) \rangle ]\approx \eta_n \langle I_2 (\bar{\x}_2 )
\rangle/ \langle I_1(\x_1) \rangle $. Since the point $\x_1$ is fixed
at $P_2$ where the test intensity is nonzero, the visibility, and
hence the signal-to-noise ratio is proportional to the intensity in
the reference arm.

Conversely, by increasing the near-field coherence, the diffraction
pattern gradually appears in the autocorrelation function of the test
arm, displayed by frames (c) of Figs.~\ref{fig:z115} and
\ref{fig:z300}.  In these plots the diffraction peak values measured
via the autocorrelation (triangles) are compared to the values
measured by coherent laser illumination (circles).  For $\snf=33~\mu$m
the partial coherence of the light is already enough to permit an
almost perfect pattern reconstruction in the autocorrelation function.

The results presented in this section evidence a clear complementarity
between the ghost diffraction scheme and the HBT scheme, that will be
further discussed in the next section.

A final remark is the following:  had we used the spatial average technique, some
information on the diffraction pattern would have been preserved in
the correlation when increasing the spatial coherence. In this technique,
in fact, the pixel position $\x_1$ in the test arm is scanned together
with $\x_2$; in this way, if some information is present in the test arm
intensity distribution, this is retrieved from the correlation. 
By increasing the spatial coherence, the diffraction pattern
becomes visible in the intensity profile of the test
arm as shown by Figs.~\ref{fig:z115}(a), \ref{fig:z300}(a), and becomes 
also visible in the correlation as a function of $\x_1$. But,
obviously, as the diffraction pattern appears in the test arm, is not
possible any more to speak about "ghost diffraction".

\section{Conclusion}
\label{sec:Conclusion}

We have shown that coherent imaging with incoherent classical thermal
light is able to produce the interference pattern of a pure phase
object.  This provides the ultimate demonstration that entanglement is
not needed to do coherent imaging with incoherent light, not even in
the case of a pure phase object. As our group has pointed out in
previous publications
\cite{thermal,ferri:2004,brambilla:2004a,lugiato:2004,gatti:2005a,gatti:2006,bache:2005,bache:2005b},
the only evident advantage of using entangled light might be that of
obtaining a better visibility.

A remarkable aspect of the present experiment is the degree of
incoherence of the pseudo-thermal speckle beams used. In order to
render the beams incoherent with respect to the object (a standard
transmission grating beam splitter with 80 grooves per mm), we had to
create speckles which in the object plane had a size of $2.0~\mu$m.
This was made possible by exploiting the so-called near-field
scattering \cite{giglio:2000,giglio:2004}, in which the speckles are
created so close to the source that their size is governed solely by
the roughness of the scattering medium.

In such conditions of spatial incoherence, we have shown that no
information on the phase object is present in the light outcoming from
the object: neither the far field intensity distribution of the test
arm nor its autocorrelation function (HBT scheme) reveal the
diffraction pattern. This information is instead present in the cross
correlation between the test arm and a reference arm that never passed
through the object (ghost diffraction).  Our results indeed evidence
that, when trying to extract information on a pure phase object, there
exist a clear {\em complementarity} between the ghost diffraction
scheme and the HBT scheme. In the HBT scheme the presence of a certain
degree of spatial coherence is the essential ingredient that permits
to extract some phase information, and the information becomes more
correct as the coherence increases. Conversely, the ghost diffraction
scheme works as a coherent imaging scheme only thanks to the spatial
{\em incoherence} of the light, and the more the light is
incoherent, the better the information is reconstructed. 
These results contradicts what was indicated in the introduction of
Ref.~\cite{abouraddy:2004}, where the possibility of doing coherent
imaging in a ghost imaging scheme employing splitted thermal light was
ascribed to the presence of spatial coherence.

\section*{Acknowledgments} 
While finalizing this manuscript we have been informed of an other
experimental observation of a phase object \cite{gori:2005} by using
split thermal light.  This work was carried out in the framework of
the FET project QUANTIM of the EU, of the PRIN project of MIUR
"Theoretical study of novel devices based on quantum entanglement",
and of the INTAS project "Non-classical light in quantum imaging and
continuous variable quantum channels". M.B. acknowledges financial
support from the Carlsberg Foundation as well as The Danish Natural
Science Research Council (FNU, grant no.  21-04-0506).

\bibliographystyle{c:/LocalTexMf/prsty}
\bibliography{d:/Projects/Bibtex/literature}

\begin{thebibliography}{10}

\bibitem{klyshko:1988}
D.~N. Klyshko, Zh. Eksp. Teor. Fiz. {\bf 94},  82  (1988), [Sov. Phys. JETP
  {\bf 67}, 1131-1135 (1988)].

\bibitem{belinskii:1994}
A.~V. Belinskii and D.~N. Klyshko, Zh. Eksp. Teor. Fiz. {\bf 105},  487
  (1994), [Sov. Phys. JETP {\bf 78}, 259-262 (1994)].

\bibitem{strekalov:1995}
D.~V. Strekalov, A.~V. Sergienko, D.~N. Klyshko, and Y.~H. Shih, Phys. Rev.
  Lett. {\bf 74},  3600  (1995).

\bibitem{pittman:1995}
T.~B. Pittman, Y.~H. Shih, D.~V. Strekalov, and A.~V. Sergienko, Phys. Rev. A
  {\bf 52},  R3429  (1995).

\bibitem{ribeiro:1994}
P.~H. {Souto Ribeiro}, S. Padua, J.~C. {Machado da Silva}, and G.~A. Barbosa,
  Phys. Rev. A {\bf 49},  4176  (1994).

\bibitem{ribeiro:1999}
P.~H. {Souto Ribeiro}, S. Padua, and C.~H. Monken, Phys. Rev. A {\bf 60},  5074
   (1999).

\bibitem{saleh:2000}
B.~E.~A. Saleh, A.~F. Abouraddy, A.~V. Sergienko, and M.~C. Teich, Phys. Rev. A
  {\bf 62},  043816  (2000).

\bibitem{abouraddy:prl-josab}
A. F. Abouraddy, B. E. A. Saleh, A. V. Sergienko, and M. C. Teich, Phys. Rev.
  Lett. {\bf 87}, 123602 (2001); J. Opt. Soc. Am. B {\bf 19}, 1174 (2002).

\bibitem{abouraddy:2004}
A.~F. Abouraddy, P.~R. Stone, A.~V. Sergienko, B.~E.~A. Saleh, and M.~C. Teich,
  Phys. Rev. Lett. {\bf 93},  213903  (2004).

\bibitem{bennink:2002a}
R.~S. Bennink, S.~J. Bentley, and R.~W. Boyd, Phys. Rev. Lett. {\bf 89},
  113601  (2002).

\bibitem{bennink:2004}
R.~S. Bennink, S.~J. Bentley, R.~W. Boyd, and J.~C. Howell, Phys. Rev. Lett.
  {\bf 92},  033601  (2004).

\bibitem{gatti:2003}
A. Gatti, E. Brambilla, and L.~A. Lugiato, Phys. Rev. Lett. {\bf 90},  133603
  (2003).

\bibitem{thermal}
A. Gatti, E. Brambilla, M. Bache, and L. A. Lugiato, Phys. Rev. Lett. {\bf 93},
  093602 (2004), quant-ph/0307187; Phys. Rev. A {\bf 70}, 013802 (2004),
  quant-ph/0405056.

\bibitem{gatti:2004a}
A. Gatti, E. Brambilla, and L.~A. Lugiato,  in {\em Quantum Communications and
  Quantum Imaging}, Vol.~5161 of {\em Proc. of SPIE}, edited by R.~E. Meyers
  and Y. Shih (SPIE, Bellingham, WA, 2004), p.\ 192.

\bibitem{bache:2004}
M. Bache, E. Brambilla, A. Gatti, and L.~A. Lugiato, Phys. Rev. A {\bf 70},
  023823  (2004), quant-ph/0402160.

\bibitem{bache:2004a}
M. Bache, E. Brambilla, A. Gatti, and L.~A. Lugiato, Opt. Express {\bf 12},
  6067  (2004), quant-ph/0409215.

\bibitem{ferri:2004}
F. Ferri, D. Magatti, A. Gatti, M. Bache, E. Brambilla, and L.~A. Lugiato,
  Phys. Rev. Lett. {\bf 94},  183602  (2005), quant-ph/0408021.

\bibitem{brambilla:2004a}
E. Brambilla, A. Gatti, M. Bache, and L. Lugiato, Fortschr. Phys. {\bf 52},
  1080  (2004).

\bibitem{bache:2005}
M. Bache, A. Gatti, E. Brambilla, D. Magatti, F. Ferri, and L. Lugiato, Comment
  on "Entangled-Photon Imaging of a Pure Phase Object", unpublished,
  quant-ph/0504081.

\bibitem{lugiato:2004}
L.~A. Lugiato, A. Gatti, E. Brambilla, and M. Bache,  in {\em Fluctuations and
  Noise in Photonics and Quantum Optics II}, Vol.~5468 of {\em Proc. of SPIE},
  edited by P. Heszler, D. Abbott, J.~R. {Gea-Banacloche}, and P.~R. Hemmer
  (SPIE, Bellingham, WA, 2004), p.\ 262.

\bibitem{gatti:2005a}
A. Gatti, E. Brambilla, M. Bache, and L.~A. Lugiato, Laser Physics {\bf 15},
  176  (2005).

\bibitem{bache:2005b}
M. Bache, L. Lugiato, A. Gatti, and E. Brambilla,  in {\em Proceedings for II
  International Conference ``Frontiers of Nonlinear Physics''}, edited by A.
  Livtak (IAP RAS, Nizhny Novgorod, Russia, 2005), p.\ 80.

\bibitem{cheng:2004}
J. Cheng and S. Han, Phys. Rev. Lett. {\bf 92},  093903  (2004).

\bibitem{wang:2004}
K. Wang and D.-Z. Cao, Phys. Rev. A {\bf 70},  041801R  (2004),
  quant-ph/0404078.

\bibitem{cai:2004d}
Y. Cai and S.-Y. Zhu, Opt. Lett. {\bf 29},  2716  (2004).

\bibitem{valencia:2004}
A. Valencia, G. Scarcelli, M. {D'Angelo}, and Y. Shih, Phys. Rev. Lett. {\bf
  94},  063601  (2005), quant-ph/0408001.

\bibitem{gatti:2006}
A. Gatti, M. Bache, D. Magatti, E. Brambilla, F. Ferri, and L.~A. Lugiato, J.
  Mod. Opt. {\bf 53},  739  (2006), quant-ph/0504082.

\bibitem{zhai:2005}
Y.-H. Zhai, X.-H. Chen, D. Zhang, and L.-A. Wu, Phys. Rev. A {\bf 72},  043805
  (2005).

\bibitem{hanburybrown:1956}
R. Hanbury-Brown and R.~Q. Twiss, Nature (London) {\bf 177},  27  (1956).

\bibitem{giglio:2000}
M. Giglio, M. Carpineti, and A. Vailati, Phys. Rev. Lett. {\bf 85},  1416
  (2000).

\bibitem{giglio:2004}
M. Giglio, D. Brogiol, M.~A.~C. Potenza, and A. Vailati, Phys. Chem. Chem.
  Phys. {\bf 6},  1547  (2004).

\bibitem{goodman:1996}
J.~W. Goodman, {\em Introduction to Fourier optics}, 2. ed. (McGraw-Hill, New
  York, 1996).

\bibitem{goodman:1975}
J.~W. Goodman,  in {\em Laser speckle and related phenomena}, Vol.~9 of {\em
  Topics in Applied Physics}, edited by D. Dainty (Springer, Berlin, 1975), p.\
  9.

\bibitem{gori:2005}
R. Borghi, F. Gori, and M. Santarsiero, private communication  (2005).

\end{thebibliography}

\end{document}